\documentclass[aps,prl,twocolumn,superscriptaddress]{revtex4}

\usepackage[dvips]{graphicx}

\begin{document}
\title{Shaken Granular Lasers}
\author{Viola Folli}
\affiliation{ISC-CNR, UOS Sapienza, P.le Aldo Moro 5, 00185, Rome, Italy}
\affiliation{Department of Physics, University Sapienza, P.le Aldo Moro 5, 00185, Rome (IT)}
\author{Andrea Puglisi} 
\affiliation{ISC-CNR, UOS Sapienza, P.le Aldo Moro 5, 00185, Rome, Italy}
\affiliation{Department of Physics, University Sapienza, P.le Aldo Moro 5, 00185, Rome (IT)}
\author{Luca Leuzzi}
\affiliation{IPCF-CNR, UOS Roma Kerberos, P.le Aldo Moro 5, 00185, Rome, (IT)} 
\affiliation{Department of Physics, University Sapienza, P.le Aldo Moro 5, 00185, Rome (IT)}
\author{Claudio Conti}
\affiliation{Department of Physics, University Sapienza, P.le Aldo Moro 5, 00185, Rome (IT)}
\affiliation{ISC-CNR, UOS Sapienza, P.le Aldo Moro 5, 00185, Rome, Italy}

%\date{\today}
\begin{abstract}
Granular materials have been studied for decades, also driven by  industrial and technological applications. 
These very simple systems, composed by agglomerations of mesoscopic particles, are characterized, in specific regimes, by a large number of metastable states and an extreme sensitivity (e.g., in sound transmission) on the arrangement of grains; they are not substantially affected by thermal phenomena, but can be controlled by mechanical solicitations.
Laser emission from shaken granular matter is so far unexplored; here we provide experimental evidence that it can be affected and controlled by the status of motion of the granular, we also find that competitive random lasers can be observed. 
We hence demonstrate the potentialities of gravity affected moving disordered materials for optical applications, and open the road to a variety of
  novel interdisciplinary investigations, involving modern statistical mechanics and disordered photonics.
\end{abstract}
\maketitle
\noindent {\it Introduction ---}
In random lasers (RLs) stimulated emission is achieved by \emph{disorder-induced} light scattering \cite{Wiersma95,Wiersma08,cao2000,Cao03r,
Balachandran:97,Lawandy94,Kaiser08, Conti08, Leuzzi09,Andreasen2011,Leonetti2011}, as observed in colloidal systems, composed by small particles suspended in thermal equilibrium in a solution, or in materials exhibiting a fixed disorder, achieved, e.g., by nano-fabrication.  RLs in shaken grains were not reported.
Granular materials (sands, powders, seeds, cements, etc.) \cite{jaeger92, jaeger96} are an extensively
studied branch of statistical mechanics, with several
important applications in chemistry or engineering.
These systems are not affected by temperature, 
and are mostly dominated by dynamical effects, while being one of the paradigms
of the statistical mechanics of disordered systems and 
still lacking general and universal theoretical descriptions.
Granular gases \cite{PoschelBook,PoschelBook2}, i.e., massive particles in rapid movement with inelastic collisions,
are obtained by putting grains under mechanical oscillation. 
By a driving solicitation, 
a gravity-sedimented ensemble of grains switches, above a critical mechanical energy, from a solid-like state to a gaseous one, whose essential feature
is the strong enhancement of fluctuations and the non-equilibrium character \cite{McLennanmBook, Puglisi2005}: even in such a dilute
configuration, regions with high density may appear.
Such a state can only be maintained by continuously furnishing mechanical energy.
\\\noindent This circumstance may have relevant implications when considering random lasing in shaken granulars, which happens when 
energy is furnished to the system not only mechanically, but also optically, by employing a light-emitting active medium. The specific and characteristic 
arrangements of the shaken grains not only can alter the RL features, but, as we demonstrate in this work, in the gaseous-like phase, may lead to the occurrence of competing RL emissions, which can be controlled by acting on the external mechanical solicitation.
Such a situation is not achievable in formerly considered RL:
 in the fixed disorder case \cite{Cao03r}, the structure cannot be externally changed; while
in the colloidal RL \cite{Lawandy94} the considered dielectric nano-particles, 
\cite{Goeuedard:93,Briskina96,Cao99,Ryzhkov05} are too light to exhibit a switchable granular behavior. 
Various authors also investigated RL by metallic nano-particles \cite{Dice05, Popov06, Meng08, Meng09}, 
however, so far, only particles with diameters of tens of nanometers were considered, which do not exhibit granular behavior because are not substantially affected by gravity. In these systems, thermal equilibrium largely limits the observable fluctuations with respect to out-of-equilibrium granulars.
%%%%%%%%%%%%%%%%%%%%%%% FIGURE 1
\begin{figure}
 \includegraphics[width=8.6cm]{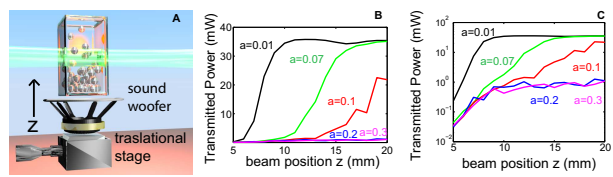}
 \caption{Experimental setup:
(A) Sketch of the granular sample placed on a vibrating woofer and a vertical translational stage, the light beam is also shown;
(B) transmission of $1064$nm continuous-wave beam for different amplitudes $a$ of the driving force;
above a critical amplitude the transmission is independent
on the value of $a$ for an increasing range in $z$, 
denoting the transition to a gaseous region; (C) as in (B) with
vertical log scale.
 \label{figure_setup}}
\end{figure}
%%%%%%%%%%%%%%%%%%%%%%% END FIGURE 1
\\\noindent Here we consider a gravity-affected granular system composed by metallic grains with millimeter size, able to macroscopically change its structural features when the status of motion is altered. We study how the dynamic structural phases affect the RL emission.
We find that an alteration of the state of motion of the grains forming the disordered laser cavity 
dramatically changes the RL emission, and sustains competitive forms of RL.
\\\noindent {\it Setup ---}
Our sample is composed by about $1500$ amagnetic-steel spherical metallic grains  with diameter of $1$~mm dispersed in a liquid RhodamineB solution 
(whose fluorescence display a broad peak around $590$~nm).
As sketched in Figure~\ref{figure_setup}A, the sample is put on a vertically ($z-$direction) vibrating plate, driven by a sound woofer,
oscillating with amplitude $a$ chosen in the normalized range $[0,1]$ and calibrated by an accelerometer ($a=0.1$ corresponds to an acceleration of $15g$ and oscillation amplitude of $0.76$mm). All the structure is placed on a vertical motorized $25$~mm translational stage.
\\We first consider the trasmission of a continuous wave  (CW) laser to determine the critical oscillation amplitude $a$ for the gaseous phase:
for any value of $a$, we make a vertical scan of the sample. 
This allows to determine the CW transmitted power at any vibrational regime and versus the position of the input laser
 (height $z$ from the bottom of the sample). The measured transmission is averaged over several temporal periods of the driving sinusoidal signal. 
As shown in Fig.\ref{figure_setup}B, for small $a$, the transmission makes an abrupt changes versus $z$, as the grains are deposited in the bottom,
and the sample transmits only when the beam is above the grains.
When the amplitude is greater than a threshold value $a\cong0.1$, the transmission profile becomes independent on $a$ and $z$;
note that the region around $z=5$~mm is affected by the transverse beam size (beam waist $6$~mm).
This change in the trend of the transmission 
signals the transition to the gaseous state, as the grains in the shaken regime have sufficient energy to uniformly explore the whole available volume,
and correspondingly intercept the beam and lower the time-averaged transmission. 
The CW beam hence simply allows to determine the onset of the gaseous state, and the corresponding critical value for $a.$
%%%%%%%%%%%%%%%%%%%%%%%%%%%
\begin{figure}
\includegraphics[width=8.6cm]{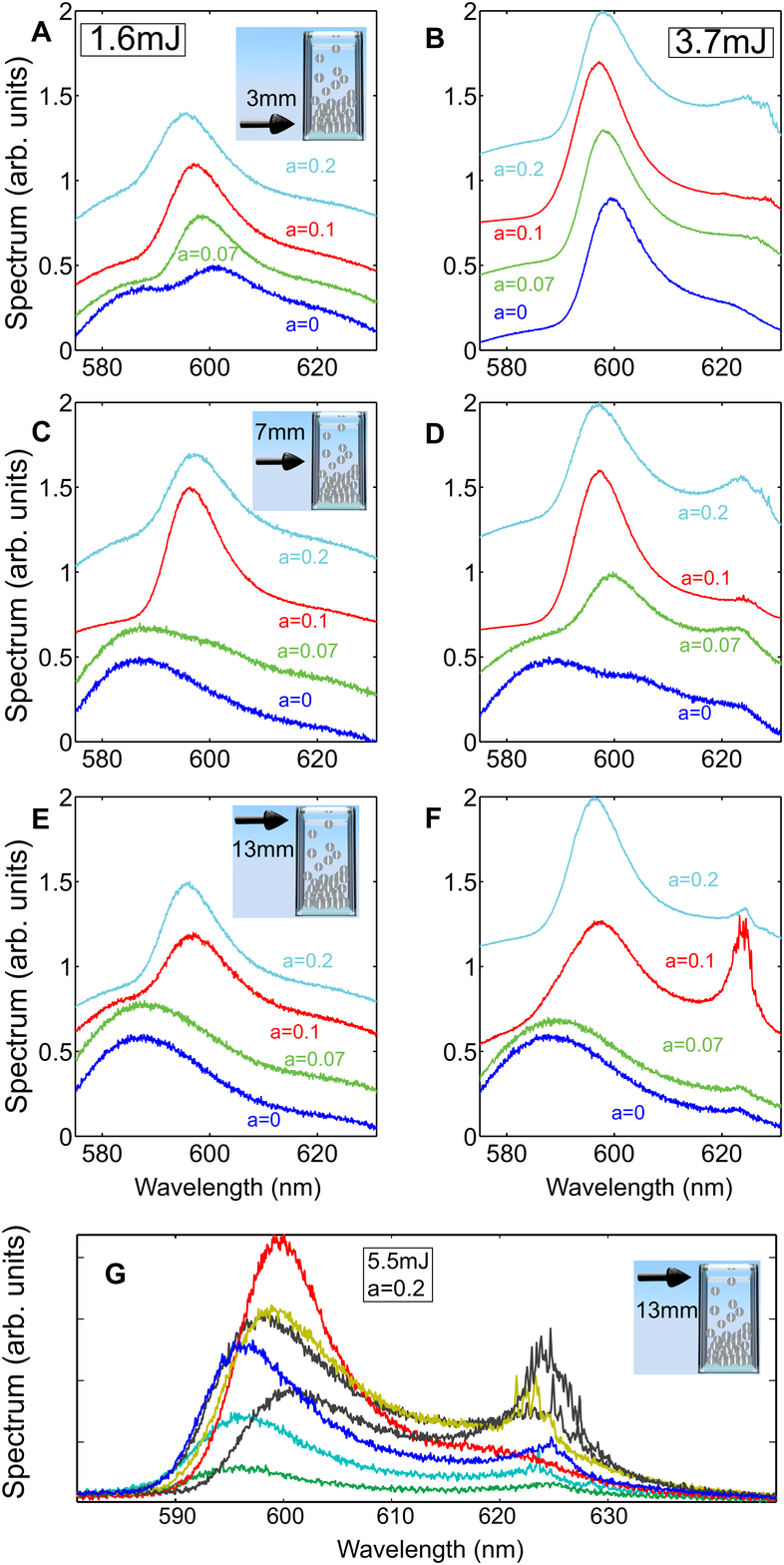}
\caption{Spectra of shaken granular lasers.
(A-F) Emission at pumping energy of 1.6mJ (left column), and 3.7mJ (right column), 
for the bottom region ($z=3$mm, panels A,B), the central region ($z=7$~mm, panels CD) and the top region ($z=13$mm, panels E,F). As indicated in the legend,
the blue line corresponds to a=0, green line to a=0.07, red line to a=0.1, sky-blue line to a=0.2; (G) single shot spectra at energy $5.5$mJ, 
$z=13$mm liquid region, and $a=0.2$.
The insets show a sketch of the sample and the arrow indicates the position of the beam with respect to the bottom. The spectra are arbitrarily shifted in the vertical direction.
 \label{figure_spectra}}
\end{figure}
%%%%%%%%%%%%%%%%%%%%%%%%%%%
\\\noindent {\it Shaken granular laser ---}
We then use an high energy pump beam ($532$~nm Nd:YAG, $10$~Hz repetition rate and $7$~ns pulse duration) 
to study RL spectra in the granular dynamic phases.
We fix the vertical position $z$ of the $532$~nm pump with sample at rest ($a=0$) and then vary the input pump energy $\mathcal{E}$
and the vibrational amplitude $a$. The interest here is to show the way spectra are affected by the mechanical energy furnished to the
granular by the sound woofer. We show in Fig.\ref{figure_spectra}A-F, spectra obtained with 
exposure time of $1$s, corresponding to $10$ single-shot average spectra;
single shot spectra are shown in Fig.\ref{figure_spectra}G.

\noindent We start considering the bottom layers of granular ($z=3$ mm). As shown in Fig.\ref{figure_spectra}A, at the pumping energy $\mathcal{E}=1.6$mJ, RL is above threshold, a peak is visible around 600 nm, which gets narrower when increasing the pump energy 
(compare panel A with panel B) \cite{Lawandy94, Conti08}. The peak is strongly affected by the amplitude $a$ (note that for $a=0$ a peak at $600$nm is visible, which is not present in the fluorescence of the Rhodamine solutions, shown, e.g., in \ref{figure_spectra}E for $a=0$, see also the discussion below).

\noindent Such a mechanism is also found at higher energies $\mathcal{E}=3.7$mJ (Fig.\ref{figure_spectra}B), however we find that above a threshold oscillation amplitude (corresponding to the onset of the gaseous phase in \ref{figure_setup}B,C), 
an additional peak aroud $620$nm appears in the spectrum (more evident when pumping in the ``central'' and ``top'' regions,
discussed below). This peak becomes more pronounced when increasing $a$.
As further detailed below, this additional RL, typical of the considered granular sample, 
competes with the other RL at $600$nm,  when the pumping optical and mechanical energies are sufficiently high.

\noindent We repeat the experiments by placing the pump beam at the edge between the grains and the liquid region of Rhodamine ($z=7$mm),
when the sample is at rest ($a=0$).
As shown in Fig.\ref{figure_spectra}C,D, the previously described phenomena are qualitatively reproduced, when changing $a$. 
However, the RL at $600$nm only appears in
the presence of shaking (which allows the grains to uniformly distribute within the pumped volume) and is more pronounced because of the higher amount of Rhodamine in the central region with respect to the bottom region discussed above.
At higher energies and higher oscillation amplitudes (Fig.\ref{figure_spectra}D), the additional peak at $\lambda=620$nm is observed in the gaseous phase. 

\begin{figure}
\includegraphics[width=8.3cm]{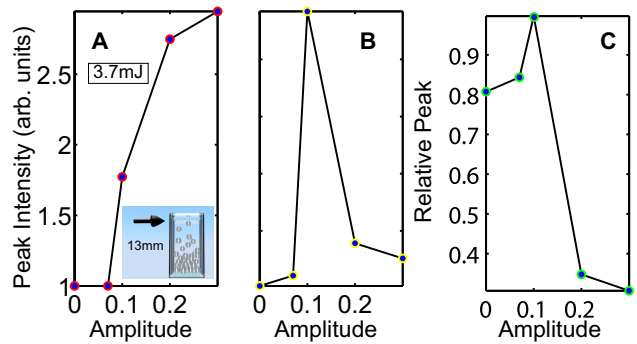}
 \caption{Competion of random lasers. In the liquid region $z=13$mm, 
peak intensity for the RL at $600$nm (A) and at $620$nm (B) versus the oscillation amplitude $a$; the ratio of peak at $620$nm and that at $600$nm is shown in panel (C).
   \label{figure_data}}
\end{figure}

\noindent When pumping in the top region ($z=13$mm) in the absence of vibration no grains are present, 
and  RL is not observed. As shown in Fig.\ref{figure_spectra}E, at energy $\mathcal{E}=1.6$mJ,
at $a=0$ and $a=0.07$, only the fluorescence of the Rhodamine is retrieved with the characteristic broad peak around $590$nm.
When the sample is put into vibration, the grains progressively fill the liquid region and the RL at $\lambda=600$nm observed.
At higher energy levels, the additional peak at $\lambda=620$~nm is more evident with respect to cases considered before. 
At sufficiently pronounced vibration, this peak is as much intense as that at $600$nm (see also Fig.\ref{figure_data});
however, at higher oscillations, the amplitude of this peak decreases again, such that an optimal amplitude exists for its observation (Fig.\ref{figure_data}D).

\noindent To address the origin of this additional peak, we show in figure Fig.\ref{figure_spectra}G a number of single shot measurements, which unveil that, at variance with the smooth broad peak at $\lambda=600$nm, also present in the single-short regime, 
the peak at $620$nm is actually composed by many narrow peaks radically changing from shot to shot, and appear smoother in figures \ref{figure_spectra}B,D,F because these correspond to averages over ten shots (exposure time $1$s). 

\noindent We interpret these additional peaks as due to optical cavities formed by few metallic grains when the sample is put under shaking because of the instantaneous arrangement of the reflecting metallic spheres. These arrangements involve few grains and change from shot to shot, hence peaks are
retrieved only for shots sustaining such specific configurations. 
\\Experiments involving a limited number of scatterers, namely segments of dye doped fibers, have shown that L\'evy flights of photons may give rise to resonant like emission \cite{sharma2005}. In our case, we are dealing with a vibrated granular system, and our results look to be another manifestation of large fluctuations often observed in granular systems,
due to both the high degree of non-equilibrium correlations and, at the same time, to their intrinsic nature of ``small systems''
(indeed, as stated above, the number of grains is of the order of $10^3$).
In this respect, our experimental results can also be interpreted as the occurrence of large amplified photon paths ("lucky photons") \cite{Mujumdar07} due to the specific density fluctuations occurring in a granular under shaking.

\noindent In Fig.\ref{figure_data}, we show the trend of the peak intensity versus $a$ for the two RL for $z=13$mm.
The peak of the RL at $600$nm (panel A of Fig.\ref{figure_data})  grows with the amplitude of oscillations. This is explained by observing that increasing the amplitude of oscillations implies increasing the interstitial holes; correspondingly, the available active medium is larger and the threshold for lasing decreases. For the RL at $620$nm, panel B shows a peak for $a=0.1$. In panel \ref{figure_data}C, the relative peak intensities are reported: for a specific oscillation $a$ the peak at $620$nm becomes more pronounced than that at $600$nm. 
%%%%%%%%%%%%%%%%
%%%%%%%%%%%%%%%%%%%%%

\noindent To investigate the coherent origin of the peak at $620$nm, we also considered a sample composed by the metallic granular system with a small concentration of dielectric ZnO particles with diameter much smaller than the grains ($d_{small}=$300nm); the results (not reported)  show that a small scattering inhibits the onset of the additional peak even in the presence of a very small concentration of the nanometric dielectric particles.

\noindent To verify that the observed localized emission is not due to the single metallic sphere, we also repeated the experiments by considering a two dimensional system, realized by a thin sample (section $1$mm$\times1$cm), such that only one layer of grain is formed, and no peak at $620$nm was observed, but only that at $600$nm (not reported). 

\noindent We hence infer that the emission at $620$nm can be ascribed to a specific three-dimensional arrangement of spheres during shaking.
This fact also allows to rule out the effect of plasmonic resonances, which are expected to play a negligible rule for the considered size of the grain ($1$~mm),
which is much greater than the wavelength. Plasmonic resonances are indeed know to be relevant for nanoparticles \cite{Dice05, Popov06, Meng08, Meng09}.
These resonances are also ruled out by the fact that the peak at $620$nm changes from shot to shot (Fig.\ref{figure_spectra}G), and hence depends on the configuration of several spheres, and not on the single sphere.
\\\noindent {\it Conclusions ---}
We have reported on the first experimental evidence of laser emission in shaken granular gases.
Our results, placed between modern photonics and statistical mechanics,
demonstrate that RL in granulars are sensitive to the specific grain distribution and
their emission can be controlled by the status of motion of the system. 
\\\noindent It is clear that a granular configuration - even in regimes of rapid shaking - is frozen at laser time-scales. Nevertheless the statistics of many emissions allow to probe those of granular configurations visited in a particular dynamical regime, which depends on shaking conditions, boundaries, grains, etc. An analytical assessment of such connection is beyond the scope of this Letter, we believe however that our observation is already sufficient to raise the case of a (statistical) connection between different granular dynamical regimes and RL: figures \ref{figure_spectra} and \ref{figure_data} provide a quantitative evidence by assessing a limited parameter region with competing RL. Configurations optimal the RL at $620$nm seem more frequent at those working points where the "density" profile (fig. \ref{figure_setup}C) is not too high and has a gradient large enough. Note that large density gradients in granular materials are typically associated with clustering and shear instabilities \cite{Zanetti93}. These instabilities are likely to support wide local fluctuations allowing the system to explore a much larger space of configurations for cavities, accessing also those able to sustain lasing effects. 
\\\noindent Our results open the way to a variety of further investigations, as RL in matter under common granular processes like compactification, metastable granular states, mixed systems, accelerated flow under gravity, supercontinuum generation, and the interaction of light with granular waves. The possibility of achieving a controlled non-equilibrium scenario in granular systems provides a variety of novel tools for the mechanical control of random photonic devices, and ultimately,  for assessing the interplay between the status of motion of mesoscopic matter and light.

\begin{acknowledgments}
\noindent The research leading to these results has received funding from the European Research Council under the European's Seventh Framework Program (FP7/2007-2013)/ERC grant agreement n. 201766, project Light and Complexity; from the Italian MIUR under the FIRB-IDEAS grant number RBID08Z9JE and under the Basic Research Investigation Fund (FIRB/2008) program grant code RBFR08M3P4. We thank MD Deen Islam for help in the development of the experimental setup, and Dr. Neda Ghofraniha for the support in the laboratory.
\end{acknowledgments}

%%% biblio
%\bibliography{MEGAbib}

\end{document}